# Graphene Dirac Fermions in 1D Inhomogenous Field Profiles: Transforming Magnetic to Electric Field

Liang Zheng Tan, Cheol-Hwan Park, and Steven G. Louie*

Department of Physics, University of California, Berkeley, California 94720 and
Materials Sciences Division, Lawrence Berkeley National Laboratory, Berkeley, California
94720

### Abstract

We show that the low-energy electronic structure of graphene under a one-dimensional inhomogeneous magnetic field can be mapped into that of graphene under an electric field or vice versa. As a direct application of this transformation, we find that the carrier velocity in graphene is *isotropically* reduced under magnetic fields periodic along one direction with zero average flux. This counter-intuitive renormalization has its origin in the pseudospin nature of graphene electronic states, and is robust against disorder. In magnetic graphene superlattices with a finite average flux, the Landau level bandwidth at high fields exhibits an unconventional behavior of decreasing with increasing strength of the average magnetic field, due to the linear energy dispersion of graphene. As another application of our transformation relation, we show that the transmission probabilities of an electron through a magnetic barrier in graphene can directly be obtained from those through an electrostatic barrier, or vice versa.

# I. INTRODUCTION

The low-energy electronic excitations of graphene are massless Dirac fermions [1], with an effective ``light speed'' of $v_0 \approx 1 \times 10^6 \text{m/s}$. Unusual phenomena associated with the Dirac Hamiltonian such as Klein tunneling [2] and the unconventional integer quantum Hall effect [3] can now be studied in bench-top graphene experiments [4-6].

A recent fruitful avenue of investigation that has brought interesting theoretical results is that of external electric [7-11] and magnetic [12-20] profiles in graphene. Such systems are also of practical interest for graphene electronics, because of effects such as electron beam supercollimation [9] in electrostatic special graphene superlattices (SGSs) and magnetic confinement of electrons in graphene [15]. Experimentally, electrostatic patterns have been fabricated on graphene with a periodicity down to 5 nm [21]; although magnetic graphene superlattices (MGSs) have not been made, techniques used in creating magnetic superlattices in 2-dimensional electron gas (2DEG) systems [22] may be relevant for this purpose. The band structure and transmission characteristics of electrostatic graphene superlattices (EGSs) on single and bi-layer graphene have been studied in Refs. [7-11] while transmission through various magnetic structures in single and bi-layer graphene were explored in Refs. 12 and13.

We demonstrate here that systems of one-dimensional (1D) electric and magnetic profiles in graphene are closely related via a transformation of the Dirac equation. This transformation has the potential to simplify the analysis of and bring new physical insights into the electronic behavior of field-induced nanoscopic and mesoscopic structures in graphene. We have made use of this transformation, together with known results for the 1D EGS, to solve for the electronic structure of a 1D MGS in the case when the average magnetic flux vanishes, $\langle B \rangle = 0$. In this case, the group velocity of the charge carriers is isotropically reduced as the strength of the magnetic field is increased (Fig. 1), a surprising result given that the external periodic magnetic field is anisotropic. The band structure for the case where $\langle B \rangle \neq 0$ is calculated using both exact numerical and perturbative methods. It is found that, in the limit of large $\langle B \rangle$, the bandwidth of the Landau bands decreases as $1/\sqrt{B}$, unlike the analogous system of a 2DEG in a periodic magnetic field where the bandwidth approaches a constant as $\langle B \rangle$ is increased [23]. We have also shown through our transformation, the relationship between the transmission probability through electrostatic and magnetic barriers in graphene.

The paper is organized as follows. In Sec. II we develop a transformation relating graphene under uni-dimensional modulated magnetic fields to the analogous system of graphene under uni-dimensional modulated electric fields. In Sec. III this transformation is applied to the magnetic graphene superlattice with $\langle B \rangle = 0$. We then examine the case where $\langle B \rangle \neq 0$. An application of the transformation to systems with finite number of magnetic barriers is presented in Sec. IV. In Sec. V, we discuss the effects of disorder on our results. We conclude in Sec. VI.

# II. THE TRANSFORMATION

**A. Dirac Hamiltonian**

We shall consider external fields with nanoscale variations much larger than the carbon-carbon distance in graphene so that intervalley scattering between the Dirac points at $K$ and $K'$ can be neglected [24,25]. We focus on low energy excitations near the $K$ point, and neglect Zeeman interactions and intrinsic spin-orbit couplings, which have respective energy scales of $\mu_B B \approx 5 \times 10^{-4}$ eV (at $B = 5$ T) and $1.7 \times 10^{-5}$ eV, according to Ref. 1. In contrast, the energy scale of an MGS is $\hbar v_0 / \sqrt{\hbar c/(eB)}$, which is at least two orders of magnitude larger than either of these two energy scales when $0.005T \leq B \leq 10T$. For simplicity, we focus on systems without large Rashba spin-orbit coupling [26-30]. Examples of such systems include graphene on Co surfaces [31-32].

We first treat the general case of electric and magnetic modulations where the field strengths vary in the $x$ direction and are constant in the $y$ direction. The electronic states of the system can be described by the Dirac equation:

$$i\hbar \frac{d\psi}{dt} = \left\{ v_0 \vec{\sigma} \cdot [-i\hbar \nabla + \frac{e}{c} \vec{A}(x)] + V(x) \right\} \psi \tag{1}$$

where the wavefunction $\psi$ is a two component spinor function, $v_0$ is the Fermi velocity of pristine graphene, and $\vec{A}(x)$ and $V(x)$ are vector and scalar potentials, respectively, which do not necessarily have to be periodic. We shall use the Landau gauge, and the magnetic field is taken to be perpendicular to the graphene layer. Writing the wavefunction as $\psi(x,y;t) = e^{ik_y y} e^{-iEt/\hbar} \varphi(x)$, the Dirac equation becomes:

$$E\varphi(x) = \left\{ -i\hbar v_0 \sigma_x \partial_x + v_0 \sigma_y [\hbar k_y + \frac{e}{c} A(x)] + V(x) \right\} \varphi(x) \tag{2}$$

**B. Complex Lorentz Boost**

Relating the electric and magnetic graphene systems is a two step process: in this subsection, we show that a complex Lorentz boost changes the Dirac equation with real magnetic (electric) fields into a Dirac equation with imaginary electric (magnetic) fields. In the next subsection, we perform an analytic continuation to relate the Dirac equation with imaginary electric (magnetic) fields to a Dirac equation with real electric (magnetic fields).

Starting from Eq. (2), we multiply throughout by $\sigma_y$ and make the unitary transformation $\varphi^{'}(x) = U\varphi(x)$ with

$$U = \frac{1}{\sqrt{2}} \begin{pmatrix} 1 & 1 \\ 1 & -1 \end{pmatrix}. \tag{3}$$

Also we transform Eq. (2) to new energy and momentum variables:

$$E = -ihv_0 k_y{}' \tag{4}$$

$$k_y = \frac{iE^{'}}{\hbar v_0}.$$

(5)

The result of these operations is to transform the original Dirac equation [Eq. (2)] into

$$E^{'}\varphi^{'}(x) = \left\{ -i\hbar v_0 \sigma_x \partial_x + \sigma_y [\hbar v_0 k_y^{'} - iV(x)] + \frac{v_0 e}{c} iA(x) \right\} \varphi^{'}(x).$$

(6)

This transformation interchanges the role of the $x$-dependent electric and magnetic fields. The transformation is actually a complex Lorentz boost with an imaginary rapidity: if a general Lorentz boost that mixes a spatial coordinate with time is represented by

$$\begin{pmatrix} t' \\ y' \end{pmatrix} = \begin{pmatrix} \cosh\theta & -\sinh\theta \\ -\sinh\theta & \cosh\theta \end{pmatrix} \begin{pmatrix} t \\ y \end{pmatrix},$$

(7)

then Eqs. (4) and (5) correspond to a Lorentz boost with rapidity $\theta = \frac{i\pi}{2}$, follow by a mirror refection $y' = -y$. It should be noted that other choices for the rapidity are possible, these will result in the mixing of the electric and magnetic profiles [33].

## C. Analytic continuation

An analytic continuation may be used to relate the solutions of the Dirac equation with imaginary fields [Eq. (6)] to a Dirac equation with real fields.

Suppose that a Dirac equation with real electric potentials (for simplicity let us assume no magnetic fields, although this can be easily added in) has been solved and the eigenenergies are known to be given by an equation $g(E, k_y, V) = 0$, where $E, k_y, V$ are all real. We argue that the Dirac equation with an imaginary electric potential of the same shape (i.e., writing $V = V_0 w(x)$ with $V_0$ now an imaginary number) has imaginary eigenenergies given by the same equation, but now with $E, k_y, V$ all imaginary.

The above argument is true since the eigenfunctions $\psi(x, y; k_y, V)$ of the original Dirac equation with real electric potential can be analytically continued to imaginary values of $k_y, V_0$. This is because the Dirac operator in Eq. (1) consists of differentiation and matrix operations, which act on the eigenfunction in the same way regardless of where $k_y$ and $V_0$ lie in the complex plane. Therefore, the analytic continuation of the eigenfunctions to imaginary $k_y$ and $V_0$ values are eigenfunctions of the Dirac equation with $k_y$ and $V_0$ imaginary. This implies that the imaginary eigenenergies are given by the same equation $g(E, k_y, V) = 0$, with $E, k_y, V(x)$ imaginary.

The system with imaginary electric fields is solved if the system with real electric fields is solved. And, by the results of II.B, the system with real magnetic fields is solved if the system with imaginary electric fields is solved. We can thus relate the solutions of graphene under electric field profiles to those of graphene under magnetic field profiles. The two steps of this transformation are summarized in Table 1. This procedure is quite general – it is applicable to inhomogeneous fields of 1D profiles of both finite and infinite extent, as well as to states with finite lifetimes (imaginary eigenenergies).

An important consideration in applying this method in practice is the fixing of boundary conditions. It is possible that after the analytic continuation, a wavefunction displays unphysical behaviour at the boundaries. Such solutions must be excluded and domain of validity of the energy dispersion relations restricted accordingly. These considerations however do not appear in the examples considered in the next section.

It should be noted that the imaginary values of energy and momentum in the intermediate stages of the transformation bear no physical significance – they are purely mathematical crutches and should not be interpreted as indicators of finite lifetimes or confined states.

## III. MAGNETIC GRAPHENE SUPERLATTICES

### A. The case with $\langle B \rangle = 0$

We now apply this method to the system of a 1D $\langle B \rangle = 0$ MGS, where $V(x) = 0$ and $A_y = A(x)$ in the Landau gauge is periodic and assumed to average to zero in one unit cell of the superlattice. Both $k_x$ and $k_y$ are good quantum numbers. The transformed system is that of a 1D EGS, (i.e. $A'(x) = 0$ and $V'(x)$ is periodic and imaginary) with $\langle \vec{E'} \rangle = 0$. We are interested in the *imaginary* $(E', k_y')$ solutions of the latter system, which we find by making use of the *real* $(E, k_y)$ solutions of the 1D EGS with $V(x)$ real. The 1D EGS with a real potential has been solved, and the energies to lowest order in $\vec{k}$ are given in Ref. 7 as:

$$E(\vec{k}) = \hbar v_0 \sqrt{k_x^2 + |f|^2 k_y^2} \qquad (8)$$

where $f = \int_{\text{unit cell}} \exp[2i \int_0^x V(x') dx' /(\hbar v_0)] \, dx$.

Using the imaginary energy eigenvalues of the 1D EGS with $k_y$ and $V_0$ imaginary via Eqs. (4) and (5), the energy bands in the MGS are found, to lowest order in $k_x$ and $k_y$, to be:

$$E(\vec{k}) = \pm \frac{\hbar v_0 \sqrt{k_x^2 + k_y^2}}{\int_{\text{unit cell}} \exp[-2 \int_0^x e A(x') dx' /(\hbar c)] \, dx} \qquad (9)$$

Remarkably, the dispersion relation near the $K$ point is isotropic, and there is *no* energy gap between the valence and conduction bands, regardless of the magnetic field strength. Furthermore, the group velocity near the $K$ point is always renormalized to be less than $v_0$, and it decreases monotonically as the strength of the magnetic field is increased. The group velocity is monotonically reduced because the derivative of the denominator of Eq. (9) with respect to $A_0$ (writing $A(x) = A_0 h(x)$ ) is non-negative, due to the fact that $\alpha(x) = 2 \int_0^x V(x^{'}) \, dx^{'} \, /(\hbar v_0)$ averaged to zero over one unit cell as shown in Ref. 7.

These results are also applicable to states around a single valley in k-space in an effective gauge field treatment of corrugated graphene [34], where a gauge field is introduced with opposite signs at each valley in order to simulate the effects of ripples in graphene. For example, applying Eq. (9) to the effective magnetic field generated by the corrugation in Fig. 2 of Ref. 26 gives a velocity renormalization that is in good agreement with the results in that figure. It should be noted that the regime considered in corrugated graphene is different from that considered here: ripples of reasonable size tend to reduce the velocity to almost zero, whereas MGSs do not.

Interestingly, carbon nanotubes under a constant, transverse magnetic field [35,36] can be considered approximately to be a special case of Eq. (9) here, for the specific value of $k_x = 0$ and $k_x = \pm 2\pi /(3L)$, corresponding to metallic and semiconducting carbon nanotubes with circumference L, respectively. In addition to corroborating the predictions of velocity reduction in metallic carbon nanotubes and gap reduction in semiconducting nanotubes in Refs. [35,36], Eq. (9) provides a description of velocities in arbitrary directions as well.

For concreteness, let us focus on the specific cases of two magnetic Kronig-Penney superlattices : i) $A(x) = A_0 \mathrm{sgn}[\sin(2\pi x / l_0)]$ which corresponds to a periodic 1D δ-function magnetic field of alternating signs, and ii) a periodic piecewise constant magnetic field of alternating sign: $B(x) = \dfrac{4 A_0}{l_0} \mathrm{sgn}[\sin(2\pi x / l_0)]$. These magnetic superlattices have period $l_0$. Evaluating Eq. (9) for the δ-function magnetic field Kronig-Penney superlattice gives

$$E(\vec{k}) = \hbar v_0 \mid \vec{k} \mid \frac{e A_0 l_0 /(4\hbar \mathrm{c})}{\sinh(e A_0 l_0 /(4\hbar \mathrm{c}))} \qquad (10)$$

This result, together with a similar formula for the piecewise constant magnetic field, is shown in Fig. 2, and the results are identical to those of numerical solutions to the Dirac equation, also shown in Fig. 2, obtained using a plane-wave basis (60 plane waves were used in the expansion of the wavefunction). In contrast, the analogous system of a 2DEG in a magnetic superlattice [37, 38] has an anisotropic energy spectrum near the ( $k_x = 0, k_y = 0$ ) point, which is expected considering the anisotropic nature of the superlattice potential.

One use of the transformation presented above is in identifying features of the EGS with features of the MGS. A simple application of Eqs. (4) and (5), shows that the isotropic velocity reduction in an MGS can be predicted from the constant (superlattice potential independent) group velocity in the $k_x$ direction in an EGS. On an intuitive level, one can think of isotropic velocity reduction as the magnetic analogue of Klein tunnelling, with both features arising from the Dirac nature (pseudospin physics) of the quasiparticles.

**B. The case with $\langle B \rangle \neq 0$**

For the case of an 1D MGS where $\langle B \rangle \neq 0$, we may write the vector potential in the Landau gauge as $A_y(x) = A_p(x) + B_0 x$, where $A_p(x)$ gives the periodic magnetic modulation and $B_0$ is the uniform background magnetic field. In this system, $k_x$ is no longer a good quantum number; we are interested in the $E$ vs. $k_y$ dispersion relation. Let us first consider the low $B_0$ semiclassical limit. We start with the energy spectrum of the 1D $\langle B \rangle = 0$ MGS found above and treat the background magnetic field as a perturbation. In this limit, the quasiparticles circulate along constant energy contours in momentum space. The quantization of these orbits leads to the formation of Landau levels. The Landau levels for pristine graphene in a uniform perpendicular magnetic field $B$ is $E_n = \text{sgn}(n)\sqrt{2e\hbar v_0^2 \, |n| \, B}$ where $n = 0, \pm 1, \pm 2, \ldots$ [5]. Since the introduction of a periodic modulating magnetic field leaves the conic energy spectrum intact and only renormalizes the group velocity, the Landau levels for the 1D $\langle B \rangle = 0$ MGS is given by the same formula as the Landau levels for pristine graphene, except for the renormalization of $v_0$. This is in agreement with the numerical solution of Eq. (1) in the low $B_0$ regime (see Fig. 3a). Since we have not assumed any range of values of $A_p$, this regime includes (at least when all the magnetic fields are small) the experimentally convenient situation of constructing the superlattice using strips of ferromagnetic material arranged in a regular spacing, which corresponds to $B_0 \approx B_p$, where $B_p$ is the periodic magnetic field. A measurement of the Landau level spacings would be one means to directly verify the isotropic velocity reduction discussed above.

The higher Landau levels are not flat (as a function of $k_y$), but show broadening in the form of oscillations as a function of $k_y$ (Fig. 3a). This behavior can be understood by considering $k_y$ as the parameter that controls the position of the wavefunctions along the $x$ direction under the gauge we adopted [1]. Changing $k_y$ changes the local environment felt by the wavefunction, and thus changes its energy. From this argument, the period of oscillations is $l_0/l_B^2$, where $l_0$ is the size of the unit cell and $l_B = \sqrt{\hbar c / (e\langle B \rangle)}$ is the magnetic length associated with the average background magnetic field strength. As can be seen in Fig. 4a, this period agrees with the results of numerical calculations (by diagonalizing the Hamiltonian in a plane-wave basis).

The lower-energy Landau levels are not affected because these states (Fig. 4b) have few nodes, and the distance between nodes is typically much larger than $l_0$ (in the limit of low $B$), so that those states effectively perform an "averaging" of the local magnetic field and their energies are not greatly affected by their position. On the other hand, higher-energy states might have a node-to-node distance comparable to $l_0$. It would then be possible to position such a state so that the peaks coincide with regions of high (or low) magnetic field, and thus affect the magnetic field strength "felt" by those states and hence their energies. This criterion for the onset of energy bands has been verified for the states in Fig. 3a.

As the strength of the background magnetic field is increased from zero, the bandwidth of the Landau level energy bands first increases monotonically from zero (not shown for the range of magnetic fields plotted in Fig. 3b), and then fluctuates , similar to the analogous system of a 2DEG in a 1D periodic magnetic modulation [23]. However, in the limit of large $B/B_l$ with $B_l$ defined as $\hbar c/(el_0)^2$, (i.e. in the limit the magnetic length $l_B = \sqrt{\hbar c/(eB)}$ becomes significantly smaller than the period $l_0$ of the superlattice), a qualitative difference between the two systems arises, in that the bandwidth approaches a constant in the case of the 2DEG, while it vanishes in the limit of large magnetic fields in the case of graphene. To obtain a physical understanding of this limit, we take the unperturbed system to be graphene in a uniform background magnetic field, while the perturbation $\Delta H = (ev_0/c)A_p(x)\sigma_y$ is the periodic modulating magnetic field. Using the zeroth order wavefunctions

$$\varphi_n(x) = \begin{pmatrix} |n-1\rangle \\ |n\rangle \end{pmatrix},$$ (11)

the first order correction to the energy is found to be

$$\Delta E = \sum_G \frac{ev_0}{c} A_G e^{-iGk_y l_B^2} \left\langle n-1 \left| e^{iGl_B(a+a^+)/\sqrt{2}} \right| n \right\rangle$$

$$\approx \sum_G \frac{ev_0}{c} A_G e^{-iGk_y l_B^2} iGl_B \sqrt{\frac{n}{2}}$$ (11)

in the limit $l_B/l_0 \ll 1$. Here, $n$ is the Landau level index, $|n\rangle$ the quantum harmonic oscillator eigenstates, $a$ and $a^+$ the creation and annihilation operators, and $A_G$ are the Fourier components of the periodic vector potential $A_p(x)$. The bandwidth falls off as $1/\sqrt{B}$ as $B \to \infty$, and the bandwidths of successive bands increases as $\sqrt{n}$. The numerical results of our calculations shown in Fig. 3 are in good agreement with these asymptotic results.

This peculiarity can be understood as a consequence of the linear dispersion relation of graphene and the fact that the energy levels of graphene in a uniform magnetic field grow as $\sqrt{B}$ rather than linearly in $B$ as in the case for a 2DEG. In the limit of large $B$, the wavefunctions are well localized, and one can consider the difference in energies between two states of the same Landau band localized at different positions in a saw-tooth type of MGS vector potential. Each of these two states is in a local environment of an (approximately) uniform magnetic field with strengths $B+B_1$ and $B+B_2$ ($B_1, B_2 \ll B$), and so the difference in their energies is approximately $\sqrt{2n(B+B_1)} - \sqrt{2n(B+B_2)} \approx (B_1-B_2)\sqrt{n/(2B)}$, which is in agreement with the result from perturbation theory.

## IV. FINITE MAGNETIC BARRIERS

In this section, we relate the transmission probability through single magnetic barriers in graphene to the transmission probability though electrostatic barriers in graphene. This is done using the complex Lorentz transformation developed in previous sections. For simplicity, we consider here square electrostatic or vector potential barriers, as shown in Fig. 5.

The transmission coefficient for a vector potential barrier such as in Fig. 5a is given by Eq. 6 of Ref. 14. This equation gives the transmission coefficient $t$ in terms of the angles of propagation inside ($\theta$) and outside ($\phi$) the vector potential barrier. The variables $\theta$ and $\phi$ are easily expressed as functions of $k_y$, $E$ and $A_0$, (the transverse momentum, the energy of the propagating wave, and the vector potential height, respectively.) Once this is done, the transmission coefficient $t(k_y, E, A_0)$ will be a function of the vector potential amplitude, the transverse momentum, and the energy. To relate this to the electrostatic barrier, Eqs. 4, 5 as well as $V' = i(v_0 e / c)A_0$ are used. The last equation $V' = i(v_0 e / c)A_0$ comes from a comparison of Eqs. 2 and 6.

If these substitutions are made in $t(k_y, E, A_0)$, an expression $t'(k_y', E', V')$ is obtained, which is the transmission coefficient through an electrostatic barrier in graphene. It can be checked with Ref. 39 that this is indeed the correct expression for the transmission coefficient. Fig. 5c,d show representative transmission probabilities (as a function of the incident angle $\phi$) for both types of barriers.

## V. DISORDER

An experimental realization of a magnetic (or electrostatic) superlattice will not be perfectly periodic due to variations in both the period of the superlattice and the strength of the local magnetic fields. We have simulated such randomness using a supercell approach, where we have solved for the bandstructure of a simulation cell consisting of 30 smaller unit cells. Each unit cell has a period which follows a normal distribution with a randomness parameter $r = \sigma / \mu$, where $\sigma$ and $\mu$ are the standard deviation and mean of the normal distribution, respectively. Similarly, the strength of the magnetic field in each unit cell is also normally distributed. We have performed calculations using values of $r$ up to 0.1. An ensemble average of 20 independent random magnetic configurations was taken in the calculations. A broadening of 0.2, in the energy units of Fig. 6, was used in the density of states calculation.

The density of states (after ensemble averaging) is shown in Fig. 6. In pristine graphene, the density of states is linear in the energy from the Dirac point energy. In the presence of a perfect magnetic superlattice, the density of states is still linear, but increased from the pristine graphene case, due to the velocity reduction effect described above. Fig. 6 shows that this observation remains true even if the magnetic superlattice is disordered. The density of states for the disordered magnetic superlattice is approximately linear, with nearly the same slope as that of the perfect magnetic superlattice. This provides evidence that the presence of low level disorder should not change significantly the magnitude of the velocity reduction effect described above.

At low energies (E<0.2 in Fig. 6,) which correspond to approximately 1/10$^{th}$ of the bandwidth of a perfect superlattice in the limit of no magnetic field, the density of states of the disordered magnetic superlattice is not linear, but instead approaches a finite value as the energy decreases to zero. For a superlattice of period L=100nm, this energy range is E < 2 meV.

## VI. CONCLUSIONS

We have discovered a transformation relating electronic properties of 1D electrostatic and magnetic structures on graphene. This transformation can be used to obtain the energy dispersion relation of one system if the energy dispersion relation of the other is known. The method is applicable to a wide range of potential profiles. The transformation relations provide a useful platform upon which other graphene nano/mesoscopic structures may be analyzed and understood. As examples of its applicability, we have analyzed both magnetic superlattices in graphene, as well as finite magnetic barriers in graphene.

We found that graphene massless Dirac fermions under magnetic profiles exhibit behaviors qualitatively different from those of the conventional 2DEG. In the magnetic graphene superlattice with no net magnetic flux, the Dirac cone displays isotropic velocity reduction, despite the anisotropic magnetic field configuration. The magnetic graphene superlattice with net magnetic flux has Landau level bandwidths that decrease with increasing average magnetic field, due to the linear energy dispersion of graphene. We have also shown that a small amount of disorder in the superlattice does not have a significant effect on these results.

## ACKNOWLEDGEMENTS

We thank Dmitry Novikov for fruitful discussions. LZT and analytic theory studies were supported by NSF Grant No.DMR07-05941 as well as the UC Berkeley Endowment Fund Fellowship for Graduate Education. CHP and numerical simulations were supported by the Director, Office of Science, Office of Basic Energy Sciences, Division of Materials Sciences and Engineering Division, U.S. Department of Energy under Contract No. DE-AC02-05CH11231. Computational resources have been provided by NERSC and TeraGrid.

*Note added.* At the final stage of writing up this manuscript, we became aware of a preprint [18] which also remarked the isotropic velocity renormalization in magnetic graphene superlattices. Both the focus of and the theoretical concepts developed in our manuscript, however, are very different from those in Ref. 18.

\* Electronic address: sglouie@berkeley.edu

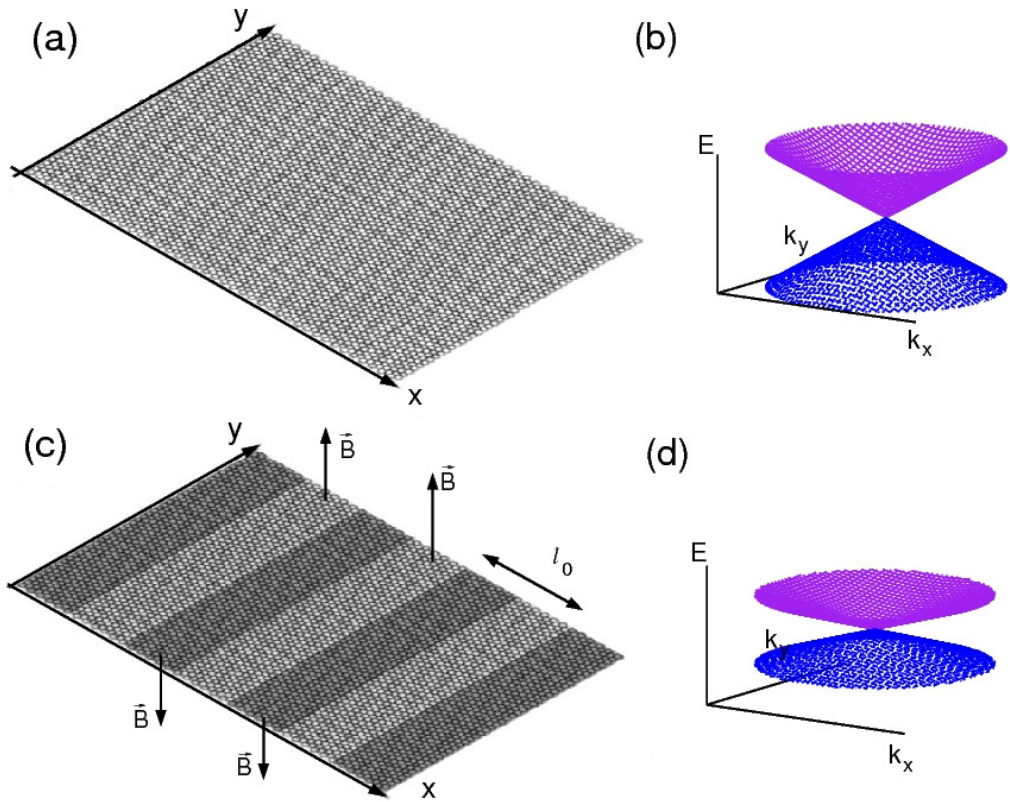

Fig 1: (a) Structure of pristine graphene. (b) Band dispersion of pristine graphene near the *K* point. (c) Structure of a 1D MGS, with the darker regions denoting a magnetic field pointing along the −z direction and lighter regions denoting a magnetic field pointing along the +z direction. This structure repeats itself in both the x and y directions. (d) The isotropically renormalized band structure of a 1D MGS of the kind shown in (c).

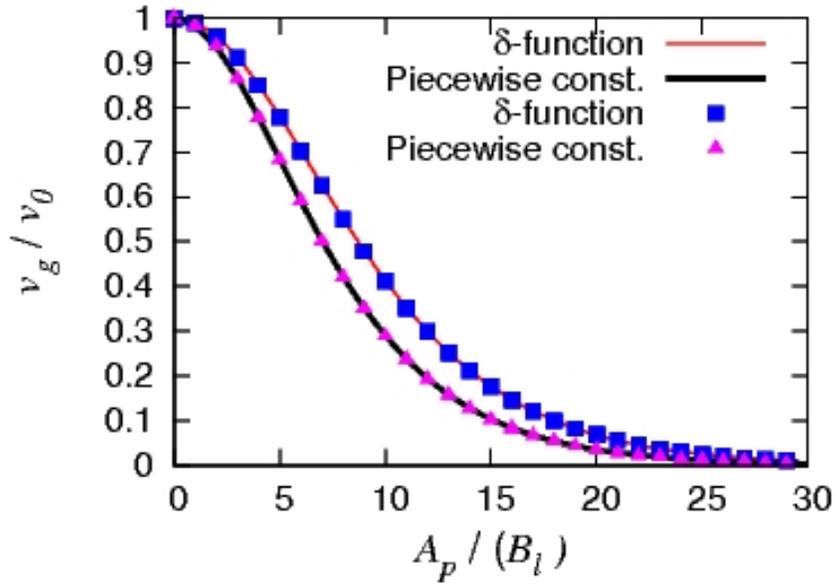

Fig. 2: The ratio of the Fermi velocity $v_g$ in the presence of a periodic magnetic field to the Fermi velocity $v_0$ of pristine graphene near the $K$ point is plotted as a function of the vector potential magnitude $A_p$, for both a δ-function magnetic field and a piecewise constant magnetic field Kronig-Penney superlattices. The analytical (lines) and numerical (symbols) results are in agreement. $l_0$ is the superlattice period and $B_l = \hbar c /(e l_0^2)$ is the characteristic magnetic field strength associated with $l_0$. The group velocity is identical in all directions. For $l_0 = 100\,\text{nm}$ and a magnetic field of $1.8\,\text{T}$, $v_0$ is renormalized by a factor of $1/2$.

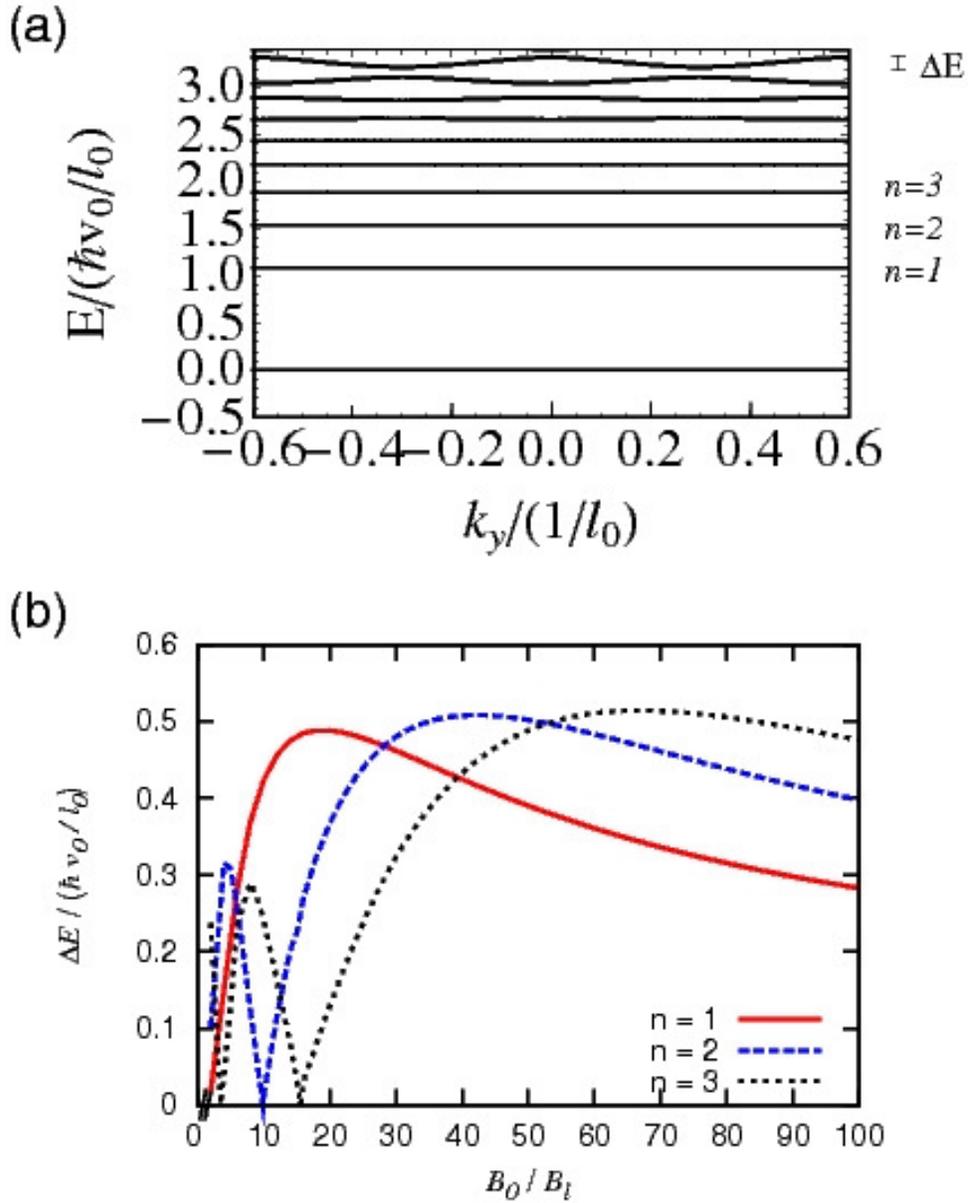

Fig. 3: (a) The energy bands for a piecewise constant magnetic field pattern with magnetic field strength $B_p / B_l = 1.4$, immersed in a uniform background field of $B_0 / B_l = 0.6$, where $B_l = \hbar c /(e l_0^2)$ is the characteristic magnetic field strength associated with the superlattice periodicity. The first 10 bands are plotted. (b) The bandwidths ΔE of the first three bands, as a function of $B_0$, of a piecewise constant magnetic field with magnetic field strength $B_p / B_l = 2$, immersed in a uniform background field of strength $B_0$.

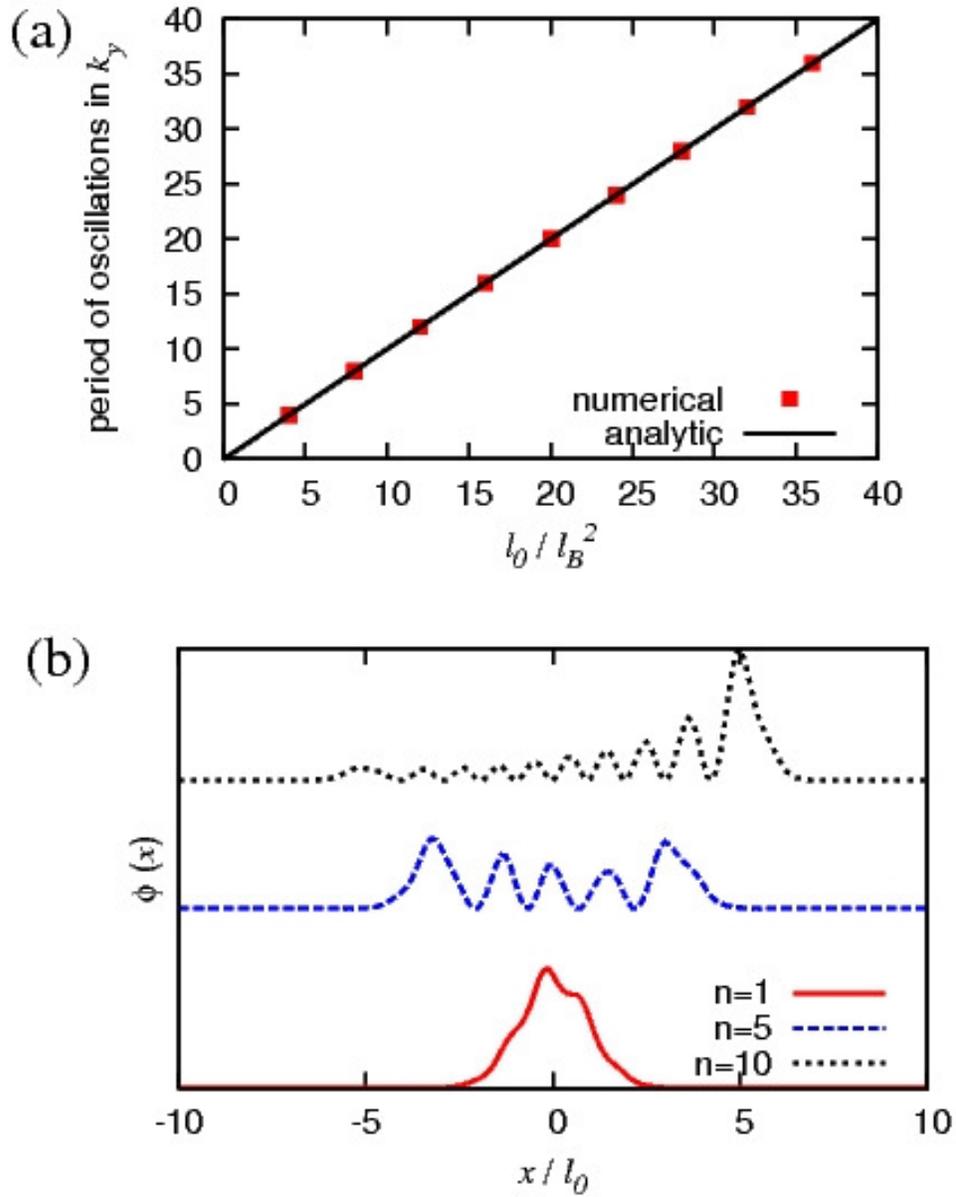

Figure 4: (a) A comparison of the period of oscillation of the Landau bands in $k_y$ as obtained from the numerical calculations with the analytic prediction that this period is equal to $l_0/l_B^2$, where $l_0$ is the size of the unit cell, and $l_B = \sqrt{\hbar c/(e\langle B\rangle)}$ is the magnetic length associated with the average background magnetic field strength. (b) Three representative wavefunctions for the system with $B_0 \neq 0$. The same parameters are used as in Fig. 3a: a piecewise constant magnetic field pattern with magnetic field strength $B_p/B_l = 1.4$, immersed in a uniform background field of $B_0/B_l = 0.6$, where $B_l = \hbar c/(el_0^2)$ is the characteristic magnetic field strength associated with the superlattice periodicity. The index n in this figure refers to the nth Landau level as defined in Fig 3a.

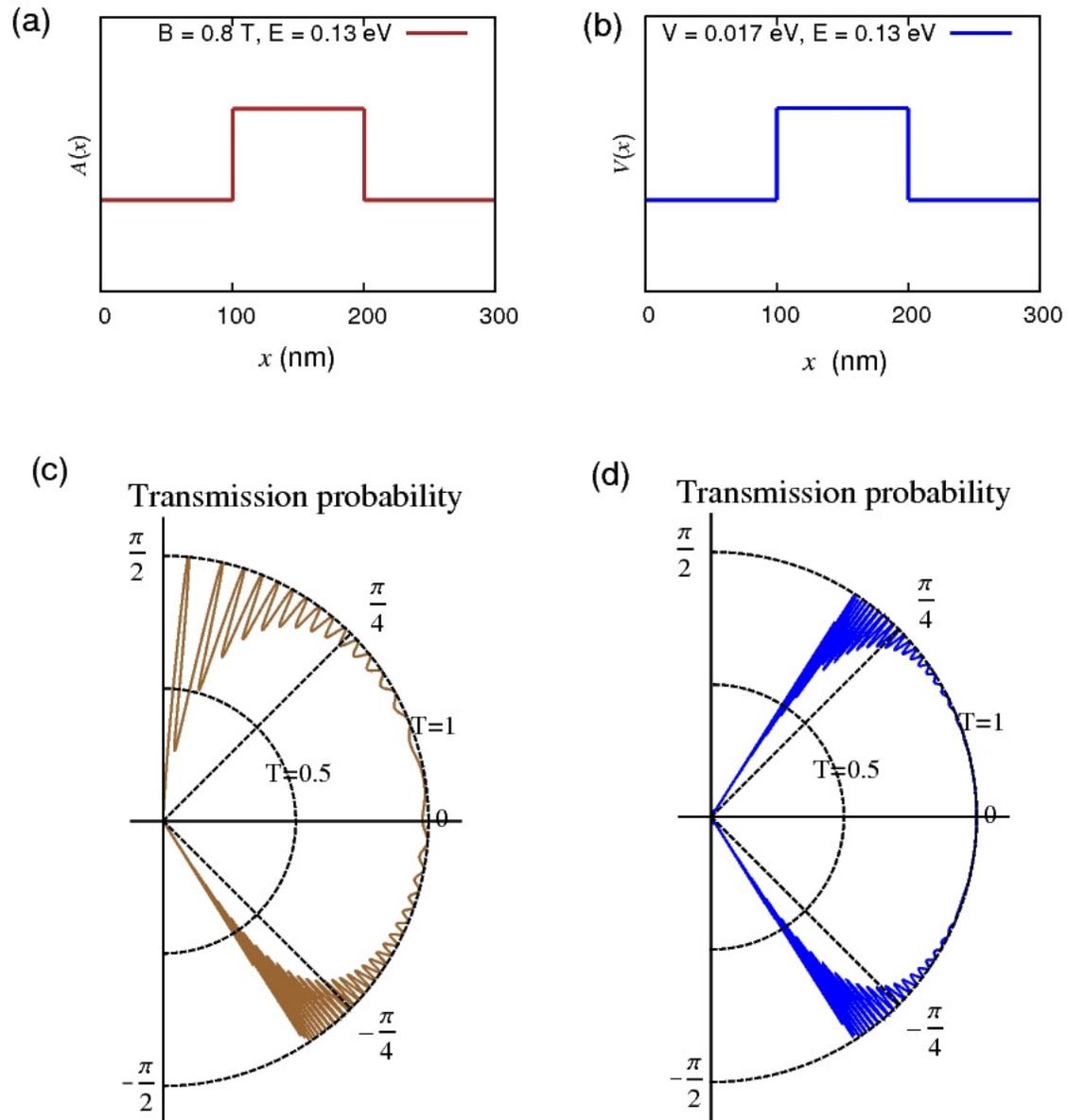

Figure 5: Vector potential (a) and electrostatic (b) barriers in graphene, and the transmission probabilities through the vector potential (c) and electrostatic (d) barriers, as a function of incident angle.

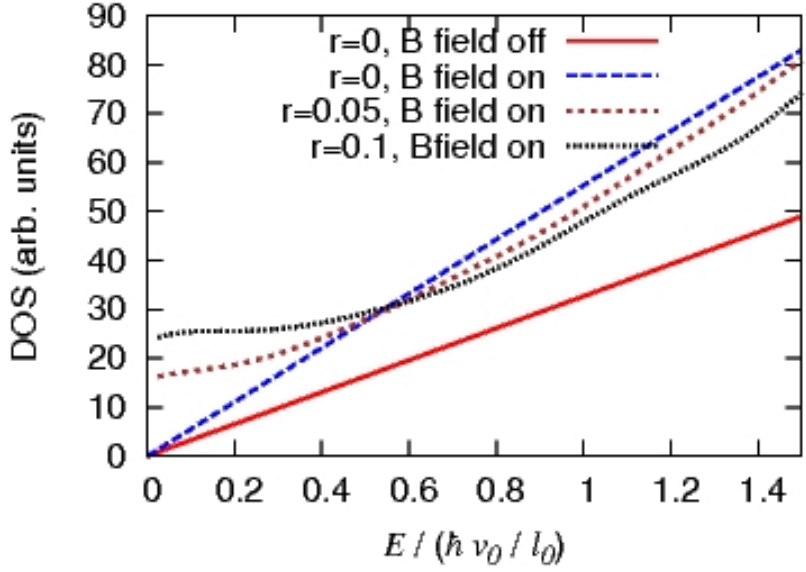

Figure 6: The ensemble-averaged density of states of a random magnetic superlattice, with randomness parameters r = 0.05 and r = 0.1 (see text), compared with a perfectly periodic superlattice (r = 0). The density of states of pristine graphene is also shown (r = 0, B field off ). The velocity reduction factor that corresponds to this change in density of states is $v/v_0 = 0.58$. The energy range in this plot is 1/2 the bandwidth of a empty-lattice graphene superlattice.

| Type of system | Dirac equation and wavefunction |
|---|---|
| Real magnetic (RM): $E, k_y, A$ real | $E\varphi^{RM}(x) = \left\{ -i\hbar v_0\sigma_x\partial_x + v_0\sigma_y\left[\hbar k_y + \dfrac{e}{c}A(x)\right]\right\}\varphi^{RM}(x)$ |
| | $\psi^{RM}(x,y;t) = \varphi^{RM}(x)e^{ik_y y}e^{-iEt/\hbar}$ |
| Imaginary electric (IE): $E, k_y, V$ imaginary | $E\varphi^{IE}(x) = \left\{ -i\hbar v_0\sigma_x\partial_x + v_0\sigma_y\hbar k_y + V(x)\right\}\varphi^{IE}(x)$ |
| | $\psi^{IE}(x,y;t) = \varphi^{IE}(x)e^{ik_y y}e^{-iEt/\hbar} = \varphi^{IE}(x)e^{\kappa y}e^{\varepsilon t/\hbar}$ , where $k_y = -i\kappa,\ E = i\varepsilon$, with $\kappa, \varepsilon$ real |
| Real electric (RE): $E, k_y, V$ real | $E\varphi^{RE}(x) = \left\{ -i\hbar v_0\sigma_x\partial_x + v_0\sigma_y\hbar k_y + V(x)\right\}\varphi^{RE}(x)$ |
| | $\psi^{RE}(x,y;t) = \varphi^{RE}(x)e^{ik_y y}e^{-iEt/\hbar}$ |
| Relation between wavefuntions | To obtain $\varphi^{IE}$ from $\varphi^{RE}$, perform an analytic continuation in $E, k_y, V_0$. |
| | To obtain $\varphi^{RM}$ from $\varphi^{IE}$, replace the imaginary V with real A, and make a unitary transformation, as described in the Eqs. 3-5. |

Table 1: Various stages of the transformation taking a magnetic structure to an electrostatic structure on graphene. The form of the wavefunctions and the corresponding Dirac equation are shown for each stage.